\documentclass[12pt]{article}
\usepackage{graphicx,epsfig,amsmath,amssymb,amsthm, mathrsfs, pstricks} 
\usepackage[left]{lineno}

\begin{document}


\title{\bf Some Consequences of the Thermodynamic Cost of System Identification}

\author{{Chris Fields}\\ \\
{23 Rue des Lavandi\`{e}res, 11160 Caunes Minervois, France}\\ 
{fieldsres@gmail.com}\\
{ORCID: 0000-0002-4812-0744}}
\maketitle

\begin{abstract}
The concept of a ``system'' is foundational to physics, but the question of how observers identify systems is seldom addressed.  Classical thermodynamics restricts observers to finite, finite-resolution observations with which to identify the systems on which ``pointer state'' measurements are to be made.  It is shown that system identification is at best approximate, even in a finite world, and that violations of the Leggett--Garg and Bell/CHSH (Clauser-Horne-Shimony-Holt) inequalities emerge naturally as requirements for successful system identification.
\end{abstract} 
~

\section{Introduction}

The idea that all finite observers are characterized by uncertainty and must pay, in energetic currency, to reduce their uncertainty was introduced into classical physics by Boltzmann \cite{boltzmann:1896}.  Shannon~\cite{shannon:48} showed that information obtained from observations can be naturally quantized into answers to yes/no questions and hence measured in bits. Landauer \cite{landauer:61, landauer:99} then showed that such information has been ``obtained'' and is available for future use only after it has been irreversibly recorded on some physical medium.  The resulting classical theory of \textit{observation}---the 
exchange of energy for information---states that, for any finite, physically implemented observer $O$, each bit of irreversibly recorded uncertainty reduction (equivalently, each bit of information gain) costs $c^{(O)} k_B T$, where $k_B$ is Boltzmann's constant, $T$ is temperature, and $c^{(O)} \geq$ ln2 is a measure of $O$'s information-aquisition efficiency that can for simplicity be considered constant.  As all classical observations in practice take place at $T > 0$, this energetic cost is always positive.  This classical theory of observation has two familiar practical consequences: observations are limited to finite resolution and records of their outcomes to finite bit strings, and only some finite number of such finite-resolution observations can be made in any finite time.

The consequences of this classical, thermodynamic limitation to finite, finite-resolution observations have been investigated in both classical and quantum settings, particularly as they bear on issues of noise (i.e., uncontrolled degrees of freedom) and measurement uncertainty.  It~has been known since the pioneering work of Spekkens and colleagues \cite{spekkens:07, bartlett:12}, for example, that classical statistical mechanics reproduces wave-packet quantum theory in the special case in which wave packets are Gaussian.  Jennings and Leifer \cite{leifer:16} review this and other work, showing that classical statistical mechanics reproduces ``quantum'' features and behavior including the uncertainty principle, non-commutativity of measurements, state teleportation and the no-cloning theorem when a finite-resource restriction limiting the number and resolution of measurements is imposed.  Krechmer~\cite{krechmer:18} shows that ``quantum'' measurement disturbance and non-commutativity of observables result whenever two measurement devices are calibrated using the same physical standard.

My aim here is to investigate a different set of consequences of the classical thermodynamic restriction to finite energy resources and hence to finite, finite-resolution observations: its consequences for an observer's ability to \textit{identify} the physical system being observed.  The question of system identification has largely been neglected by theoretical physics, although it is of obvious practical relevance to experimental physics.  Discussions of quantum measurement, for example, standardly examine the interaction between an observer and a fixed, well-defined system that is typically stipulated a priori by stipulating its Hilbert space (for reviews, see \cite{landsman:07, schloss:07}).  System identification has received more attention from engineers and computer scientists.  Moore, for example, proved in 1956 that finite, finite-resolution observations cannot fully determine the state space of an otherwise-uncharacterized physical system; in particular, they cannot determine its state-space dimension $d$ (\cite{moore:56} Theorem 2; see \cite{fields:13, fields:16} for discussion).  This result underlies the proven unsolvability of the halting problem in classical computer science \cite{hopcroft:79}.  Both of these results, however, rely on limits in which numbers of degrees of freedom become arbitrarily large, and neither considers the quantitative cost of system identification.

Here I characterize the thermodynamic cost of system identification in a general, operational framework covering both classical and quantum systems and investigate some of its consequences. The next section characterizes the system identification problem operationally as a search problem constrained by a finite-resource restriction.  The consequences of this restriction for system identification and characterization are then discussed 
 (Section~\ref{systems}), and~the finite-resource restriction is shown to forbid the arbitrary refinement of state spaces of observed systems to assumed ``objective'' or ``ontic'' state spaces even when these are (Section~\ref{refinement}).  I then focus on two types of system-identification problems that regularly arise in practice: the identification of a single system at multiple, significantly separated times (Section~\ref{leggett}) and the identification of a single system by multiple, spacelike-separated observers (Section~\ref{bell}). I~show in each case that classical correlations of measurement outcomes are insufficient, in principle, for reliable system identification. A similar point regarding the second class of problems has been made previously by Grinbaum \cite{grinbaum:17}.  Violations of Leggett--Garg \cite{emary:14} and Bell \cite{mermin:93} inequalities, respectively, thus arise naturally as requirements for reliable system identification in these settings.  These results extend and elaborate on previous work of a more systems-theoretic nature~\mbox{\cite{fields:13, fields:16, fields:12a, fields:12b}}.  The general theory of observation as a physical process, including the central role of the observer's memory as an encoding of observational outcomes, has also recently been discussed by Kupervasser~\cite{kupervasser:17}.

\section{Formalizing System Identification as a Search Process}\label{defs}

\begin{figure}
\centering
\includegraphics[width=14 cm]{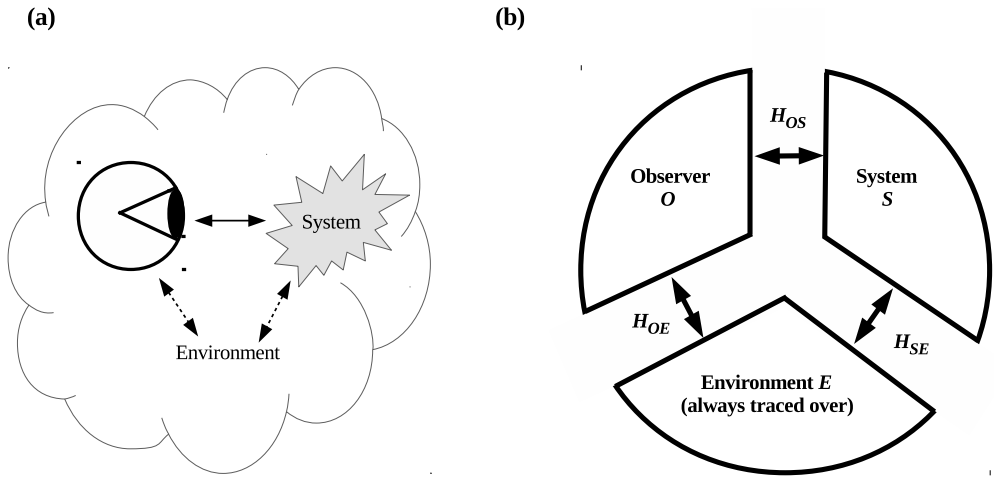}
\caption{(\textbf{a}) A classical observer interacts with a system of interest; both are embedded in a surrounding environment. (\textbf{b}) Interactions between observer ($O$), system of interest ($S$) and environment ($E$) enabling environmental decoherence.  The Hamiltonian $H_{OS}$ transfers outcome information from $S$ to $O$; $H_{SE}$,~and $H_{OE}$ decohere $S$ and $O$ respectively. Adapted from Figure 1 in ref. \cite{tegmark:12}.}
\label{fig1}
\end{figure}

Characterizing the thermodynamic cost of system identification requires redescribing observation in a way that makes the process of system identification explicit.  Consider the standard, classical ``picture'' of observation shown in Figure \ref{fig1}. Here the ``observer'' is a physical system that interacts with a ``system of interest'' to obtain observational outcomes.  Both observer and system are embedded in a surrounding environment, which can be regarded as ``everything else'' in the universe.  This classical picture of observation is carried over unchanged into quantum theory, where the ``observer'' now terminates the von Neumann chain \cite{vonNeumann:55} by recording their outcome(s) in a thermodynamically irreversible way.  It provides, by including the surrounding environment, the setting for environmental decoherence \cite{zeh:70, zeh:73, zurek:81, zurek:82, joos-zeh:85, zurek:98, zurek:03}.  Tegmark has emphasized that the observer $O$ in this setting comprises \textit{only} 
the degrees of freedom that record observational outcomes, while the system $S$ comprises only the ``pointer'' degrees of freedom that specify these outcomes; all other degrees of freedom are considered part of the ``environment'' $E$ and traced over \cite{tegmark:12}.  Tracing out the environment assures that information about the state of $S$ reaches $O$ only through the channel defined by the $O-S$ interaction, specified in Figure \ref{fig1}b by the Hamiltonian $H_{OS}$.  The alternative channel via the environment $E$, given by the Hamiltonian $H_{SE} + H_{E} + H_{OE}$, contributes only classical noise.  In the alternative ``environment as witness'' formulation of decoherence developed by Zurek and colleagues \cite{zurek:04, zurek:05, zurek:06, zurek:09}, $O$ is assumed to be located sufficiently far from $S$ that $H_{OS} \sim 0$.  In this formulation, all information about $S$ obtained by $O$ flows through the channel $H_{SE} + H_{E} + H_{OE}$.  The state $|E \rangle$ of the environment is regarded as ``encoding'' this information, with the encoding of information about the positions of macroscopic objects by the ambient photon field as the canonical example.

This conventional conception of observation, even when made precise using the formalism of decoherence, tells us nothing about how the observer \textit{identifies} the system of interest.  The system $S$ is given a priori in Figure \ref{fig1}: the interactions $H_{OS}$, $H_{OE}$, $H_{ES}$, $H_S$, and $H_E$ are all assumed to be given and well-defined.  To include system identification in the picture, it is useful to describe it in operational terms.  Suppose I want Alice to report the observational outcome registered by a particular macroscopic apparatus located in a laboratory filled with many other systems.  How much information do I need to give Alice to assure that she reports the outcome from the \textit{right} apparatus?  In this scenario, the~finite-resource restriction on Alice is clear: I can give Alice at most a finite description of the apparatus that I want her to report an outcome from.  I could instruct her, for example, to locate a black laptop labeled ``data 3,'' running linux, with a counter window open, and to report the outcome displayed in the counter window.  I could add that ``data 3'' is connected to an ADC in the third rack from the right wall.   Alice must then enter the laboratory and \textit{look for}, using observational means at her disposal, an~apparatus matching my finite description.  The informational basis of this operational scenario can be made precise as follows:

\begin{quote}
\textbf{Finite-resource restriction}: No observer can employ more that a finite number of finite-resolution observational outcomes to identify a system of interest. 
\end{quote}

Classically, an observer subject to the finite-resource restriction has only a finite number of finite-resolution criteria for system identification; in quantum theory, this corresponds to a finite number of discrete-valued observables.  Such criteria or observables can be considered to be binary without loss of generality.

\begin{figure}
\centering
\includegraphics[width=13 cm]{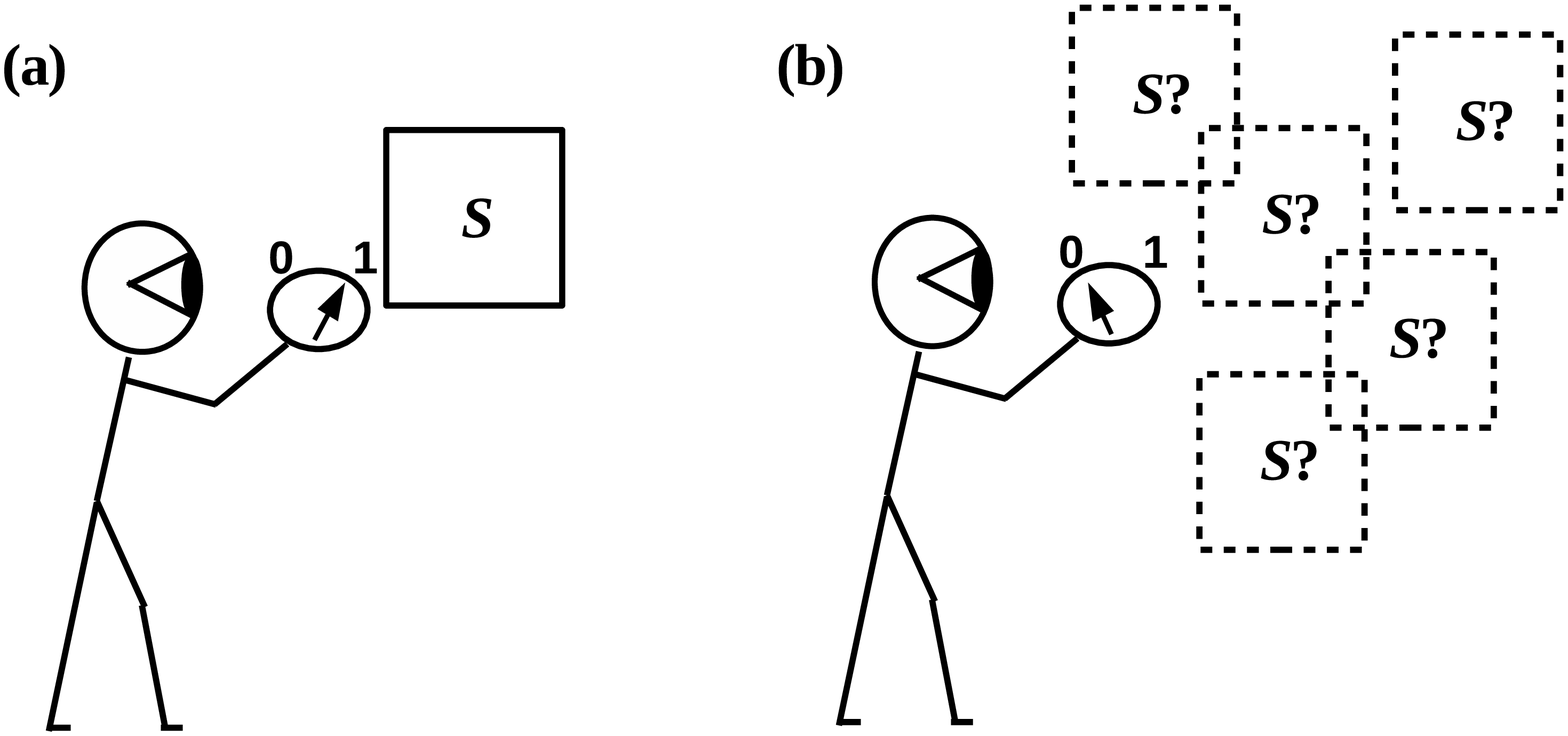}
\caption{(\textbf{a}) An observer equipped with an observable (e.g., a meter reading) interacts with a pre-given system $S$.  Adapted from Figure 1 in ref. \cite{fuchs:10}. (\textbf{b}) An observer with finite resources must \textit{look for} the system of interest by probing the ``world'' $W$ in which it is embedded.}
\label{fig2}
\end{figure}

It is obviously circular to assume that, when Alice enters the laboratory, she can identify the apparatus satisfying her finite criteria (or finite observables) without having to look at anything else: this is equivalent to assuming that the apparatus is given a priori and hence does not need to be identified.  To \textit{identify} the apparatus $S$, Alice must \textit{distinguish it}, using her criteria/observables, from everything else in the laboratory, i.e., from $E$.  Alice must, in other words, employ her criteria/observables to \textit{search} the combined system $W = SE$ until she finds $S$.   Hence, she is in the position illustrated in Figure \ref{fig2}b, not that of Figure \ref{fig2}a as is standardly assumed.

\begin{figure}
\centering
\includegraphics[width=12.5 cm]{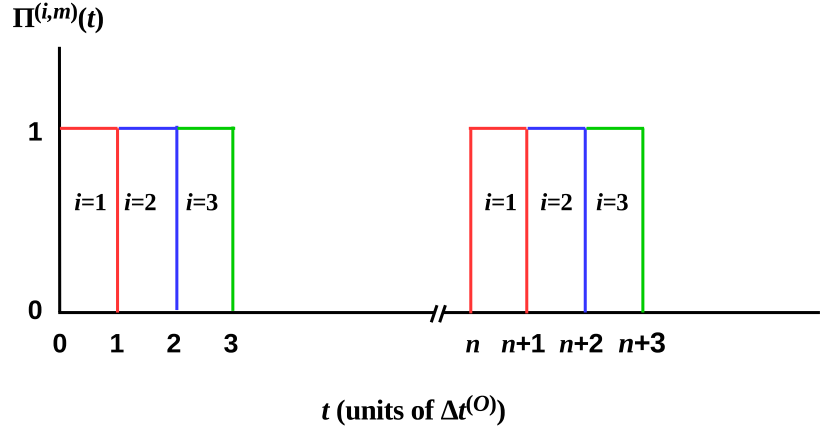}
\caption{The first three components of $\Pi^{(i,m)}(t)$ of Equation \eqref{Pi-def} in the first and $n^{th}$ cycles of deploying~the~$M_i$.}
\label{fig3}
\end{figure}

To make this idea of searching $W$ for $S$ precise, suppose as above that an observer $O$ and world $W$ are given as collections of physical degrees of freedom, and assume for the present that they are quantum systems characterized by Hilbert spaces $\mathcal{H}_{\mathit{O}}$ and $\mathcal{H}_{\mathit{W}}$, respectively.  Suppose further that $O$ can perform $n$ distinct (but not necessarily orthogonal) binary-outcome measurements $M_i$ on $W$, that $O$'s thermodynamic cost per bit of recorded outcome is $c^{(O)} k_B T$ as above, that deploying the $M_i$ has no other energetic consequences, and that $O$'s interaction with $W$ consists entirely of deploying the $M_i$.  In~this case, each of the $M_i$ can be regarded as extracting one bit of information from $W$ and exhausting $c^{(O)} k_B T$ of waste heat into $W$.  The operations $M_i$ can be regarded informally as ``questions to Nature'' such as ``is what I see before me a laptop?'' or ``is it black?'' and formally as Hermitian operators on $\mathcal{H}_{\mathit{W}}$ in the usual way.  No assumption need be made at this point about whether the $M_i$ commute; this question is addressed in Section~\ref{systems}.  For simplicity, suppose $O$ deploys the $M_i$ one at a time in the fixed order $i = 1, ..., n$, that each of the $M_i$ is deployed for a fixed time $\Delta t^{(O)}$, the time required for $O$ to record one bit, and that $O$ makes $m$ cycles of measurements.  The total elapsed time during which $O$ makes measurements on $W$ is then $nm \Delta t^{(O)}$.  Taking the $O-W$ interaction to be given by a Hamiltonian operator $H_{OW}$ on $\mathcal{H}_{\mathit{O}} \otimes \mathcal{H}_{\mathit{W}}$, the total action is

\begin{equation}
\int_{t=0}^{nm \Delta t^{(O)}} dt ~H_{OW}(t) = nm \Delta t^{(O)} c^{(O)} k_B T.
\end{equation}
\noindent
To make 
 $H_{OW}(t)$ explicit in the simplest case of sequential, equal-duration measurements, let~$\Pi^{(i,m)}(t)$ be the rectangular Pi function with offset $i$, $0 \leq i \leq n-1$, duty cycle $n$, and the number of cycles $m$, i.e.,

\begin{equation}
\Pi^{(i,m)}(t) = \sum_{j=0}^{m-1} \Pi (t -(nj + i + 1/2) \Delta t^{(O)})  \label{Pi-def}
\end{equation}
\noindent
where

\begin{equation*}
\Pi (t) = \left\{
\begin{array}{rl}
0 & \text{if } |t| > 1/2\\
1/2 & \text{if } |t| = 1/2\\
1 & \text{if } |t| < 1/2
\end{array} \right..
\end{equation*}
\noindent
This $\Pi^{(i,m)}(t)$ is a sequence, starting at $t = i$, of $m$ unit-height rectangular pulses with width $\Delta t^{(O)}$ and separation $n \Delta t^{(O)}$ as shown in Figure \ref{fig3}.  In this case, we can write, for $0 \leq t \leq nm \Delta t^{(O)}$,

\begin{equation} \label{box-HOW}
H_{OW}(t) = \sum_{i=0}^{n-1} \Pi^{(i,m)}(t) M_i,
\end{equation}
\noindent
with the heat dissipated by the action of the $k$th measurement operator during the first $j \leq m$ cycles of measurement given by

\begin{equation}
(1/ \Delta t^{(O)}) \int_{t=0}^{nj \Delta t^{(O)}} dt ~\Pi^{(k,j)}(t) M_k = j c^{(O)} k_B T.
\end{equation}
\noindent

If the 
requirement of a fixed sequence of equal-duration measurements is now dropped and $O$ is simply assumed to make $N$ total observations, Equation \eqref{box-HOW} can be generalized, for $0 \leq t \leq N \Delta t^{(O)}$, to

\begin{equation} \label{gen-HOW}
H_{OW} (t) = \sum_{i=1}^n \alpha_i (t) M_i,
\end{equation}
\noindent
subject to the constraints that, at all $t$,

\begin{equation}
\sum_{i=1}^n \alpha_i (t) = 1,
\end{equation}
\noindent
and, for any positive integer $k < N$,

\begin{equation}
(1/ \Delta t^{(O)}) \sum_{i=1}^n \int_{t = k \Delta t^{(O)}}^{(k+1) \Delta t^{(O)}} dt ~\alpha_i (t) M_i = c^{(O)} k_B T. \label{diss}
\end{equation}
\noindent
Here the function $\alpha_i (t)$ is naturally interpreted as the probability of deploying the measurement $M_i$ at $t$.  The sequence of outcomes obtained will depend on the $\alpha_i (t)$; however, the incremental heat dissipation, expressed in Equation \eqref{diss}, of the measurements will not.

 \vspace{-4pt}

With $\Delta t^{(O)}$ finite, $t$ can be treated as having only integer values $k \Delta t^{(O)}$ and hence regarded as a counter.  This counter must be internal to $O$, as otherwise the values of $t$ would be observational outcomes obtained from an external clock by some subset of the $M_i$ and the above representation would be circular.  The record of $O$'s $N$ observations can, in this case, be represented as Table \ref{table1} indexed by integer values of $t$:

\begin{table}
\caption{Sample record of $O$'s observational outcomes from $W$, starting at $t = 1$.}
\centering
\label{table1}
\begin{tabular}{ccc}
\hline
\textbf{Step}	\textbf{\textit{t}}& \textbf{Measure} $M_i, i \leq n$	& \textbf{Outcome} $x_i \in \{0, 1\}$\\
\hline
1		& 1			& 1\\
2		& 2			& 0\\
3       & 2         & 1\\
$...$     & $...$       & $...$\\  
$N$       & 4         & 0\\
\hline
\end{tabular}
\end{table}
\noindent

A table of this form contains all of the information about $W$ available to $O$ following $N$ observations.  The energetic cost of these data to $O$ is $N c^{(O)} k_B T$, which is dissipated into $W$ as waste heat.  The counter $t$ can, alternatively, be regarded as counting sets of $k$ simultaneously measurable outcomes obtained ``in parallel'' at a cost of $k c^{(O)} k_B T$; however, here we will maintain the convention that outcomes are obtained sequentially at discrete time steps.

As noted above, finite observations at finite resolution cannot fully determine the state space of an otherwise-uncharacterized system \cite{moore:56}.  The only information about $W$ available to $O$ are the outcomes $x_1 ... x_N$ of $N$ finite-resolution observations; hence, $O$ cannot determine the state space of $W$, i.e., the~assumed Hilbert space $\mathcal{H}_{\mathit{W}}$ or even its dimension $d_W$, and \textit{ipso facto} can specify the measurements $M_i$ being performed on $W$ only operationally.  The complete set of possible outcomes of the $M_i$ are, however, fully specified: each action with $M_i$ produces an outcome $x_i \in \{0, 1 \}$.  Associating each of these $x_i$ with a unit basis vector $\vec{i}$ constructs a binary space $\mathcal{W}$ with dimension $d_{\mathcal{W}} \leq n$ (equality if the $M_i$ are orthogonal and all are employed at least once), which we can call the \textit{apparent} or \textit{observable state space} of $W$ for $O$.  Each ``observation'' by $O$ can, therefore, be thought of not as an action with some $M_i$ on $W$ but as an operation on $\mathcal{W}$ with a binary-valued POVM $E_i$ that selects the same outcome $x_i$ as $M_i$.  The Hilbert spaces standardly employed in quantum theory are constructed in this way using possible outcomes as basis vectors and are hence ``apparent'' in this sense.  The operators $M_i$ are, similarly, standardly defined in terms of the outcomes they produce, i.e., as operators on such apparent state spaces; in this case, the relation $E_i = M_i^{\dag}M_i$ can be viewed as operationally defining $M_i$.  This standard practice justifies our starting assumption that $O$ and $W$ can be treated as quantum systems.  The same formalism can be employed to represent finite, finite-resolution measurements of classical systems by requiring that all states be Gaussian \cite{bartlett:12,leifer:16}.

\section{Distinguishing Reference from Pointer Degrees of Freedom} \label{systems}

We now turn to the question of commutativity requirements for the $M_i$.  To be of empirical interest, a ``system'' $S$ must   (1) be 
 distinguishable from its surroundings, (2)~be sufficiently persistent in time to permit multiple observations (at minimum, ``preparation'' followed by observation), and~(3)~occupy more than one state.  Determining the state of $S$ at multiple times requires an ability to distinguish $S$ from its surroundings, i.e., to \textit{identify} $S$, at multiple times.  Hence, any system $S$ of empirical interest can be decomposed as $S = PR$, where the generalized ``pointer'' component $P$ indicates the system's time-varying state, and the remaining ``reference'' component $R$ permits, by remaining in a time-invariant state $| R \rangle$, re-identification of $S$ at multiple times.  For ordinary items of laboratory apparatus like voltmeters or oscilloscopes, size, shape, mass, and the layout of controls and displays on the surface are components of $R$ and their fixed, system-identifying values are components of $| R \rangle$, while the position of the apparatus, what the leads are connected to, control settings, and what is indicated on the displays are components of $P$.  The state $|S \rangle$ of $S$ is then given by $|S \rangle = |R \rangle |P \rangle$ with $|R \rangle$ fixed and only $|P \rangle$ free to vary.  Requiring $S$ to be identifiable by observation is thus requiring $|S \rangle$ to be separable as $|R \rangle |P \rangle$.  If my laptop's mass or the color of its exterior casing, for example, become entangled with what is displayed in one of its windows, I will no longer be able to identify it by~observation.

This requirement of re-identifiability can be formulated using Zurek's notion of a ``predictability sieve'' \cite{zurek:03}, a criterion that allows the future state of a system, here the state $| R \rangle$ of the time-invariant reference component $R$, to be predicted with confidence.  Predictability is only assured if, for all $i$,
\begin{equation}
[H_W + H_{OW}, M^{(R)}_i] = 0 \label{commute1}
\end{equation}
where the measurement operators $M^{(R)}_i$ act on $R$ but not $P$ (cf. \cite{zurek:03}; Equation 4.41).  In practice, it is sufficient that, for all $i$, $[H_W + H_{OW}, M^{(R)}_i] < \delta$ for some sufficiently small $\delta$ over the course of an experiment involving multiple observations.  Given Equation \eqref{gen-HOW}, the predictability sieve condition expressed in Equation \eqref{commute1} requires that system identification using the $M^{(R)}_i$ does not disturb system identity and that pointer-state measurements using some set of measurement operators $M^{(P)}_j$ that act only on $P$ do not disrupt system identification,~i.e.,
\begin{equation}
[M^{(R)}_i, M^{(R)}_j] = 0 \quad \mathrm{and} \quad [M^{(R)}_i, M^{(P)}_j] = 0 \label{commute2}
\end{equation}
\noindent
for all $i, j$.  Nothing, however, requires the pointer measurements $M^{(P)}_j$ to all mutually commute, and they do not, for example, if calibration is included (\cite{krechmer:18} or Section~\ref{leggett} below).  With these definitions, system \textit{identification} is distinct from system \textit{preparation}; operations employed for preparation must preserve system identity and thus   must commute with the $M^{(R)}_i$, but need not, and in general will not, commute with the $M^{(P)}_j$.  Preparation and observation of the ``pointer state'' $| P\rangle$ of $P$ will be considered equivalent in what follows.

In terms of the equivalent operators $E_i$ defined on the apparent state space $\mathcal{W}$, an ``observable system'' $S$ in $W$ can now be operationally defined as

\textbf{Definition:}
An \textit{observable system} $S$ in $W$ is a collection $(E^{(R)}_i, x^{(R)}_i)$ of $1 < k < n$ mutually commuting POVMs $E^{(R)}_i$ defined on the apparent state space $\mathcal{W}$ with specified outcomes $x_i \in \{0, 1\}$ that measure ``reference'' degrees of freedom of $W$ that are fixed and no longer free in $S$ and hence ``identify'' $S$, together with a distinct collection of $1 < l < (n - k)$ POVMs $E^{(P)}_j$ defined on $\mathcal{W}$ with unspecified binary outcomes $x^{(P)}_j$ that measure ``pointer'' degrees of freedom of $W$ that remain free in $S$, where for each $E^{(P)}_j$, $[E^{(P)}_j, E^{(R)}_i] = 0$ for every~$E^{(R)}_i$.
\noindent

Note that, while $O$ and $W$ are collections of degrees of freedom and are hence ``ontic'' entities, observable systems are collections of operations and outcomes and are hence in some sense ``epistemic'' entities.  The notations ``$S$,'' ``$R$,'' ``$P$,'' and, below, ``$E$'' will be maintained for consistency with the literature, and to recognize that in practice systems are standardly defined in terms of observational outcomes as noted~above. 

The $l$ pointer degrees of freedom of $S$ comprise its pointer $P$ and their measured outcome values constitute its pointer states $|P \rangle = |x^{(P)}_1 ... x^{(P)}_l \rangle$.  While the $E^{(P)}_j$ selecting pointer outcomes are not required to commute, at least pairs of pointer outcomes must be compatible in any ``interesting'' system (an EPR/Bell experiment, for example, requires simultaneous measurement of two pointer states, the ``measurement setting'' and the outcome, by each observer (Section~\ref{bell})).  The collection $(E^{(R)}_i, x^{(R)}_i)$ of $k$ specified (POVM, outcome) pairs specifies the pointer-state independent reference component $R$ and its time-invariant state $|R\rangle = |x^{(R)}_1 ... x^{(R)}_k \rangle$.  We require that \mbox{$P \cap R = \emptyset$ and $PR = S$}.  For~macroscopic systems such as laboratory apparatus, the number of pointer degrees of freedom $l << n$.  Hence, the number of reference degrees of freedom \mbox{$k \sim n$}; this~will be assumed in what~follows.

Two observable systems $S$ and $S'$ are discernible in isolation only if they differ by at least one reference (POVM, outcome) pair.  Every observable system $S$ has a \textit{complement} $\bar{S}$ that is the maximal observable system that does not overlap $S$.  In the limit $n \rightarrow \infty$, $S\bar{S} \rightarrow W$, i.e., $\bar{S} \rightarrow E$ as defined above.  This limit cannot, clearly, be reached with finite observational resources; the consequences of this are considered in Section \ref{refinement} below.

Several remarks are in order:

\begin{enumerate}

\item
Observable systems are defined here in terms of both the assumed partition of ``the universe'' into $O$ and $W$ and the operations employed by $O$ to identify them.  They are, therefore, observer-relative in the sense defined by Rovelli \cite{rovelli:96} for quantum states.  However, as noted above, the present considerations apply to both classical and quantum systems provided the finite-resource restriction is respected.  This observer-relativity naturally suggests counterfactual indefiniteness, i.e., that ``unidentified systems have no states'' (cf. \cite{peres:78}), regardless of the equations of motion they obey while being observed.

\item
The ``world'' $W$ is not an observable system.  As $S \rightarrow W$ the notion of ``system identification'' loses any operational meaning.

\item
The apparent state space $\mathcal{W}$ coarse-grains $W$.  As will be made precise in the next section, unless $d_W >> n$ (hence effectively, $W >> O$), waste heat cannot be dissipated by $O$ and commutativity of observables breaks down.  This corresponds to the ``large environment'' assumption of~decoherence.

\item
No assumption is made that $W$ exhibits objective classical randomness.  The characterization of the energetic cost of observation as waste heat reflects $O$'s objective uncertainty about the distribution of this energetic input across the degrees of freedom of $W$.

\item
The requirement that every $E^{(P)}_j$ commutes with all $E^{(R)}_i$ enables repeated pointer measurements to have the same outcome, and hence enables ``ideal measurements'' as defined by Cabello \cite{cabello:18}, provided calibration procedures are implemented as discussed in Section~\ref{leggett}.

\item
The support of the $E^{(P)}_j$ and $E^{(R)}_i$ in $\mathcal{W}$ can be considered the apparent or observable state space $\mathcal{S}$ of $S$; again, this is the usual approach to defining state spaces for stipulated quantum systems.  State transitions in $\mathcal{S}$ can be represented as actions of a discrete \textit{observed propagator} $\mathcal{P}^S: |S \rangle|_{t} \mapsto |S \rangle|_{t + 1}$.  This $\mathcal{P}^S$ maps each observational outcome to its successor  and so can be regarded as defining a computational process, regardless of whether the system $S$ is classical or quantum, provided the finite-resource restriction is respected \cite{fields:12b}.
\end{enumerate}

\section{System Identification Cannot Be Arbitrarily Refined}\label{refinement}

In contrast to the operational, observer-dependent conception of ``systems'' defined above, classical (or ``effectively classical'') macroscopic systems such as laboratory apparatus are standardly thought of as both observation- and observer-independent.  They are, in particular, standardly viewed both as invariant under decompositions of ``the universe'' into  
 alternative observer---world pairs---and as well-defined independently of any particular observer or observables (see \cite{bell:90} for an example of this ``realist'' position).  Let us use the notation $\mathbb{S}$ to indicate an observer-independent (``objective'' or ``ontic'') system, i.e., one that is considered well-defined in the absence of any observers, reserving $S$ for ``observed systems'' defined operationally as above in terms of sets of observational outcomes.  It~is, for~example, completely standard in classical physics to describe two observers interacting with or otherwise obtaining information about a single, observer-independent, macroscopic ``object'' $\mathbb{S}$.  This assumption of observer-independence is often carried over into quantum theory.  Extensions of the environment as witness formulation of decoherence to models of quantum Darwinism \cite{zurek:06,zurek:09} or quantum-state broadcasting \cite{chiribella:06,korbicz:14}, for example, postulate that multiple observers can independently interact with separable, redundant encodings of the eigenvalues of a single, observer-independent interaction $H_{\mathbb{SE}}$ between an observer-independent quantum system $\mathbb{S}$ and its observer-independent environment $\mathbb{E}$.  That such an encoding is redundant, i.e., that the multiple ``copies'' of the information are encoded by the single, objectively well-defined interaction $H_{\mathbb{SE}}$ must be assumed a priori, as~it cannot be established by observation \cite{fields:10}.  It is also commonly assumed, for example, in stating the Pusey--Barrett--Rudolph theorem \cite{pusey:12} that multiple ``copies'' of a single quantum system can be acted upon (e.g., prepared and/or measured) independently by multiple, mutually distant observers.  The~copies in this case are assumed to objectively have all and only the same degrees of freedom, the~same self-Hamiltonian, and the same interaction with their respective environments. 

Here we consider whether, and to what extent, observers subject to the finite resource restriction imposed in Section~\ref{defs} can identify, and hence either prepare or measure, a postulated ``objective'' system $\mathbb{S}$.  We first consider, in this section, the case of a single observer $O$ interacting during one time period with a single $\mathbb{S}$.  We then consider two cases of practical interest: in Section~\ref{leggett}, that of a single observer interacting with $\mathbb{S}$ during multiple time periods and, in Section~\ref{bell},  that of multiple observers interacting with $\mathbb{S}$ during a single time period.  We show that violations of Leggett--Garg and Bell inequalities, respectively, can be interpreted as criteria for successful system identification in these two scenarios.

As noted above, the methods developed here apply equally to both classical and quantum systems provided the finite-resource restriction is respected.  Let us now assume, as is typical in classical physics and as the simplest case, that $W$ has an \textit{observer-independent, classical} state $|W\rangle$, and first consider the finite case in which $W$ can be described by a $d_W$-dimensional, classical, binary state space, e.g., a~real Hilbert space.  Let us also assume that an observer-independent, classical system $\mathbb{S}$ is embedded in $W$, that $O$ obtains information specifically from $\mathbb{S}$ while dumping waste heat specifically into an observer-independent environment $\mathbb{E}$ defined by $\mathbb{S}\mathbb{E} = W$, and that the dimension $d_{\mathbb{S}} << d_{\mathbb{E}}$.  We~assume that $O$ interacts with $\mathbb{S}$ via a set of operators $M_i$ as defined by Equation \eqref{gen-HOW} above.  The outcomes $x_i$ of this interaction can be associated with unit vectors to construct the apparent state space $\mathcal{S}$ of $O$'s observed system $S$ as described above.  In this case, $O$ can, given a sufficient number (i.e., $n \geq d_\mathbb{S}$) of binary measurement operators, refine the observed $S$ to the objective $\mathbb{S}$,i.e., the dimension \mbox{$d_{\mathcal{S}} \rightarrow d_{\mathbb{S}} << d_W$,} at~a energetic cost of

\begin{equation}
H^{(\mathbb{S})}_{diss} = (1 / \Delta t^{(O)}) \int_{t = 0}^{\tau} H_{O \mathbb{S}} dt \rightarrow d_{\mathbb{S}} c^{(O)} k_B T
\end{equation}
where $t$ is a time coordinate associated with $W$, and $\tau \rightarrow d_{\mathbb{S}} \Delta t^{(O)}$ is the interval in $t$ required by $O$ to identify $S$ at the given refinement.  By dissipating $H^{(\mathbb{S})}_{diss}$ exclusively into $\mathbb{E}$, $O$ assures that $\mathbb{S}$ remains undisturbed.   It is this transfer of waste heat to a large, unobserved, observer-independent environment that enables the typical classical assumption of arbitrary measurement resolution and hence real-valued measurement outcomes.

If the assumption that $O$ obtains information specifically from $\mathbb{S}$ is now dropped and $O$ is required to \textit{identify} $\mathbb{S}$ by observation as described above, $O$ must search and therefore interact with, in the limit, all of $W$.  In this case, refining the observed $S$ to the objective $\mathbb{S}$ requires refining the apparent state space $\mathcal{W}$ to the full ``ontic'' state space of $W$.  The number of measurement operators required is now $n \geq d_W$, and the energetic cost is now

\begin{equation}
H^{(W)}_{diss} = (1 / \Delta t^{(O)}) \int_{t = 0}^{\tau} H_{OW} dt \rightarrow d_W c^{(O)} k_B T \label{diss-limit}
\end{equation}
where now $\tau \rightarrow d_W \Delta t^{(O)}$.  In this limit, $c^{(O)} k_B T$ is transferred, on average, to every binary degree of freedom of $W$.  The environment $\mathbb{E}$ can no longer be treated as an unobserved ``sink'' for waste heat, as~in the limit every degree of freedom of $W$ must be examined to see whether it is a degree of freedom of the as-yet unidentified $\mathbb{S}$.  Equation \eqref{diss-limit} does not depend in any way on $W$ being classical  but rather is a straightforward consequence of Equation \eqref{gen-HOW}; it is, therefore, completely general.  Hence, we have the following.

\textbf{Theorem 1:}  An observed system $S$ cannot be refined to an objective system $\mathbb{S}$ with finite resources.

\textbf{Proof:}
Consider the states $|W\rangle|_{t = 1}$ and $|W\rangle|_{t = k \Delta t^{(O)}}$ acted on by measurement operators $M^{(R)}_1$ and $M^{(R)}_k$, respectively, for some $k >> 1$.  As $d_{\mathcal{W}} \rightarrow d_W$, under increasing refinement, the maximum value of $k \rightarrow d_W$, and the energy difference between $|W\rangle|_{t = 1}$ and $|W\rangle|_{t = k \Delta t^{(O)}}$ at maximum $k$, $\Delta H_{1,k} \rightarrow H^{(W)}_{diss}$.  None of the $M_i$ are, however, orthogonal to $H_{OW}$, so in this case $[M^{(R)}_1, M^{(R)}_k] \neq 0$.  This violates the predictability sieve condition expressed in Equation \eqref{commute2}, rendering $|R\rangle$ no longer invariant.  Hence, $S$ is, by definition, unidentifiable in this limit, and the desired refinement of $S$ to $\mathbb{S}$ fails.  $\square$
\noindent 

Note that Equation \eqref{diss-limit} is independent of $d_{\mathbb{S}}$: the energy dissipation required for system identification increases with $d_W$ even if $d_{\mathbb{S}} << d_W$.  As $d_W \rightarrow \infty$ or becomes continuous, arbitrary refinement of $\mathcal{W}$ requires $H_{OW} \sim H_W$ and again commutativity of the (now infinitely or continuously many) $M^{(R)}$ fails.  Theorem 1 thus provides a quantitative extension of Moore's qualitative result that finite, finite-resolution observations cannot fully determine the state space of an otherwise-uncharacterized system \cite{moore:56}, and shows that it holds even in a finite ``world'' $W$.

System identification cannot, therefore, be arbitrarily refined to the limit of an ``objective system'' even in classical physics.  The predictability sieve expressed in Equation \eqref{commute1} that allows system identification is only operable provided the measurement interaction $H_{OW} << H_W$ and the apparent state space dimension \mbox{$d_{\mathcal{W}} << d_W$}.  Coarse-graining $W$ is, therefore, required to identify any embedded system $S$, even if $W$ is classical; if observer-independent ``objective systems'' exist in $W$, identifiable systems only approximate them.  An observed $S$ can, at best, only be associated with a set $\{ \mathbb{S} \}$ of objective systems that could, in~some theoretical model specifying some set of reference operators $\{ M_i^{(R)} \}$, generate the observational outcomes $\{ x_i^{(R)} \}$ that identify $S$.  The dimensions of the elements of $\{ \mathbb{S} \}$ are constrained only by $d_W$ and $d_S$ as upper and lower bounds, respectively.  Hence, Theorem 1 rules out any confirmation by finite observations that two independently observed systems $S$ and $S'$, whether classical or quantum, are~copies of a single objective $\mathbb{S}$.

In practice, observers search for systems only locally, effectively coupling a small, searched region of $W$ to a large, unobserved reservoir---the 
rest of $W$---into which energy can be dissipated.  If~this coupling is weak and the dissipation constant $c^{(O)} >> 1$, the predictability sieve condition expressed in Equation \eqref{commute2} fails as search resolution increases, i.e. as $S \rightarrow \mathbb{S}$
.  Observers typically search even for macroscopic systems at low resolution and then refine the search slightly after plausible candidates have been identified.  One may, for~example, locate multiple systems of the right size and shape to be one's laptop and then refine the search by looking for identifying marks, checking the splash screen, etc.  Refining the search toward an ``objective'' limit by examining every transistor, much less every atom, disrupts the commutativity of the $M^R_i$ and is therefore infeasible.

\section{System Identification at Multiple Times} \label{leggett}

Let $S|_t$ be the observed system identified when $O$ deploys $n$ measurement operators $M_i$ during the interval between $(t - n \Delta t^{(O)})$ and $t$.  Theorem~1 above show that $S$ cannot be refined to some specific objective $\mathbb{S}$.  However, $S|_t$ can be associated with a set $\{ \mathbb{S} \} |_t$ of all objective systems for which the $M_i$ would yield, at $t$, the outcomes obtained.  For example, if $S|_t$ is identified by the two criteria of being red and having no linear dimension greater than 1 m, then the set $\{ \mathbb{S} \} |_t$ contains all objective systems meeting these criteria at $t$.   If $O$ deploys the $M_i$ at multiple times, a sequence $S|_t$, $S|_{t'}$, $S|_{t''}$, etc. is obtained, with corresponding sets of objective systems $\{ \mathbb{S} \} |_t$, $\{ \mathbb{S} \} |_{t'}$, $\{ \mathbb{S} \} |_{t''}$, etc.  The sequence $S|_t$, $S|_{t'}$, and $S|_{t''}$ identifies a single observed system $S$ if there is a time-invariant set of reference outcome values $\{ x^{(R)}_i \}$ that fixes a reference state $|R\rangle$ and hence a reference component $R \subset S$.  However, $O$~cannot determine by observation that $\{ \mathbb{S} \} |_t ~=~ \{ \mathbb{S} \} |_{t'}$ or even $\{ \mathbb{S} \} |_t ~\cup~ \{ \mathbb{S} \} |_{t'} ~\neq~ \emptyset$, as doing so requires determining the self-Hamiltonians of the $\mathbb{S}$, i.e., arbitrarily accurate refinement forbidden by Theorem~1.  To continue the previous example, $O$ cannot determine that each red thing will remain red or that each small thing will remain small without examining, for each thing, more degrees of freedom than color and size.  Hence, even perfect correlation of each of the reference outcome values $x^{(R)}_i$ between all pairs of measurement times cannot gaurantee that $O$ is interacting with the same objective system(s) at $t$, $t'$, $t''$, etc.  If the probability distributions over pointer outcome values $x^{(P)}_j$ are time-invariant, their time correlations are similarly insufficient to guarantee that $O$ is interacting with the same objective system(s) at all measurement times.  Hence, we have the following.

\textbf{Theorem 2:}  
If for a set of measurements $M_i$ and measurement times $t_j$ and $t_k$, the two-time outcome correlation functions $C_{jk} = \langle x_i(t_j), x_i(t_k) \rangle$ satisfy the Leggett--Garg inequality, the observed system $S$ identified by the $M_i$ cannot be associated with any single element of the set $\{ \mathbb{S} \}$ of objective systems associated with $S$.

\textbf{Proof:}
Mapping each binary outcome from $\{0, 1 \}$ to $\{ -1, 1 \}$, the Leggett--Garg inequality can be written $C_{21} + C_{32} - C_{31} \leq 1$ for consecutive measurements at $t_1, t_2$, and $t_3$ \cite{emary:14}.  The reference outcomes $x^{(R)}_i$ and hence the reference state $|R\rangle$ must remain fixed at all observation times to identify $S$; hence, the $x^{(R)}_i$ satisfy this inequality trivially.  To see that the fixed $x^{(R)}_i$ cannot identify any particular element of the set $\{ \mathbb{S} \}$ of objective systems associated with $S$, it is enough to note that obtaining $x^{(R)}_i$ from a measurement on $\mathbb{S}$ at $t$ provides no evidence that $\mathbb{S}$ was in state $|R\rangle$ at $t-1$.  Hence, if $\mathbb{S}$ is to be identified, it must be identified by correlations between the pointer outcomes $x^{(P)}_j$.  If these satisfy the Leggett--Garg inequality, however, each measurement of the $x^{(P)}_j$ is independent of all previous as well as all future measurements.  Hence, no measurement of the $x^{(P)}_j$ on $\mathbb{S}$ at $t$ can provide information about the state of $\mathbb{S}$ at $t-1$.  It is, therefore, consistent with both constant $x^{(R)}_i (t)$ and classically correlated $x^{(P)}_j (t)$ that outcomes have been obtained from a different element of $\{ \mathbb{S} \}$ at each measurement~time.  $\square$
\noindent

Theorem~2 restates, in effect, the general principle that classical correlation does not imply joint causation; even perfectly correlated outcome values can have different causal sources.  It shows, in the present context, that an observed system $S$ cannot be associated with a particular objective $\mathbb{S}$ without violating the Leggett--Garg inequality.  Violations of this inequality provide, therefore, \textit{evidence} that a single objective system has been identified over time.

To assure violations of the Leggett--Garg inequality, $O$ must choose pointer measurement operators $M^{(P)}_i$ such that, for $t_j < t_k$, $Prob(x^{(P)}_i(t_k) = 1 | x^{(P)}_i(t_j) = 1) \neq Prob(x^{(P)}_i(t_k) = 1 | x^{(P)}_i(t_j) = 0)$, i.e., the~pointer state $|P\rangle$ must ``remember'' previous applications of $M_i^{(P)}$.  Pointer states with this property are commonplace in classical systems; magnetic hysteresis and work hardening in metals are familiar examples.  Direct measurements of such states are not non-disturbing.  If the pointer states of a macroscopic apparatus ``remember'' disturbances caused by previous measurements in this way, the~standard corrective is frequent recalibration.  Calibrating an apparatus, i.e., using measurement of a designated standard to adjust (i.e., intentionally disturb), the pointer state of the apparatus, effectively erases the memory of the previous measurement-induced disturbance.  By providing evidence that the Leggett--Garg inequality has been violated, a need for re-calibration provides evidence of previous use and hence evidence that a single objective system $\mathbb{S}$, i.e., the apparatus, has been identified.

System identification over time, therefore, requires a significant asymmetry between reference and pointer degrees of freedom.  Measurements of reference degrees of freedom must be non-disturbing in order for the reference state $|R\rangle$ to remain fixed and the observed system $S$ to be identifiable.  If $S$ is to be identified with an objective $\mathbb{S}$, however, consecutive pointer measurements cannot be non-disturbing.  Re-preparing $\mathbb{S}$ between designated, non-consecutive ``informative'' measurements, i.e., re-calibration to erase the memory of previous measurements, allows the ``informative'' measurements to be mutually non-disturbing and hence ideal.

Quantum violations of the Leggett--Garg inequality can, clearly, only be observed if the pointer state component exhibiting the violation is not re-prepared by calibration between measurements.  Observing quantum Leggett--Garg violations while maintaining a constant objective $\mathbb{S}$ requires at least one pointer state component that both exhibits memory of previous measurements and can be recalibrated between measurements.  An apparatus control setting that is re-set, and hence re-prepared, between measurements satisfies this requirement.  Here the ``standard'' to which the state of the control setting is effectively being calibrated is the observer who manipulates the setting.  Note that such control settings cannot, while preserving their function of enabling re-preparation and hence re-identification, become entangled with other components of $P$ that register the observational outcomes of interest.  As~in the case of $R$ becoming entangled with $P$ discussed in Section~\ref{systems} above, entanglement between control and outcome-registering components of $P$ can lead to system-identification failure.

\section{Joint System Identification by Multiple Observers}\label{bell}

Suppose Alice deploys measurement operators $A_i$ with outcomes $a_i(t)$ to identify and obtain pointer-state outcomes from 
an observed system $S = RP$ and Bob, who is spacelike separated from Alice at each measurement time $t$, deploys measurement operators $B_j$ with outcomes $b_j(t)$ to identify {and obtain pointer-state outcomes from} an observed system $S' = R'P'$.  Under what conditions can Alice and Bob conclude, when later comparing their separate sequences of observations, that they were observing two ``parts'' of the same objective system $\mathbb{S}$?  It is useful to consider this question from the perspective of an adversarial game; from this perspective, Alice and Bob determining that they share a single $\mathbb{S}$ is equivalent to Alice and Bob determining that they share a communication channel that cannot be, or at least has not been, manipulated by an adversary, Charlie.  Suppose $S$ and $S'$ are connected by a classical, timelike communication channel $C$, such that $SCS' = \mathbb{S}$.  Under what conditions can Alice and Bob conclude that their observations of $S$ and $S'$ are unaffected by Charlie breaching and manipulating the channel $C$?  In particular, under what conditions can Alice and Bob conclude that their observational outcomes obtained from $S$ and $S'$ are not the result of Charlie breaching $C$ and sending instructions to $S$ and $S'$ that determine the observational outcomes? This question has been extensively investigated in the guise of quantum communication security~\cite{ekert:91,gisin:07}, and the answer is well known.  Any pattern of classical correlations between $|S \rangle$ and $|S' \rangle$ can be undetectably produced by a manipulative Charlie; therefore, no pattern of classical correlations between Alice's and Bob's observational outcomes can demonstrate that the channel $C$ is secure.  Hence, we~have the following.

\textbf{Theorem 3:}  
If correlations between sequences $a_i(t)$ and $b_j(t)$ of observational outcomes obtained by spacelike-separated observers $A$ and $B$ {are consistent with a deterministic hidden-variable theory}, they cannot mutually identify a single jointly observed objective system $\mathbb{S}$.

\textbf{Proof:}
{Any pattern of} correlations between the $a_i(t)$ and the $b_j(t)$ that {is consistent with a deterministic hidden-variable theory can be implemented} by Charlie; hence, any such {pattern of} correlations is consistent with $A$ and $B$ observing separate systems, both of which are manipulated by Charlie.  $\square$
\noindent

{As the reference states $|R \rangle$ and $|R' \rangle$ remain fixed throughout Alice's and Bob's measurements to identify the observed systems $S$ and $S'$, respectively, and hence remain perfectly classically correlated, Theorem 3 effectively concerns patterns of correlations between the pointer-state outcomes $a_i^{(P)}(t)$ and the $b_j^{(P')}(t)$.  Such pointer-state correlations only permit identification of a single jointly observed objective system $\mathbb{S}$ if they are inconsistent with any deterministic hidden-variable theory.}

{Patterns of pointer-state correlations that are inconsistent with any deterministic hidden-variable theory are well-known in the special case in which Alice and Bob perform a canonical EPR/Bell type experiment.  In this case, their sets of pointer measurement operators $\{ A_i^{(P)} \}$ and $\{ B_j^{(P')} \}$, respectively, each comprise one ``control setting'' observable and two mutually noncommuting, two-valued ``outcome'' observables; the observed correlations between the ``outcome'' observables are inconsistent with any deterministic hidden-variable theory if and only if they violate at least one Bell/CHSH inequality \cite{fine:82}.  Mermin \cite{mermin:90} explicitly considers deterministic hidden variables as} ``instruction sets'' carried by particles from a central source to spacelike-separated detectors in discussing {such experiments.}  Correlations that {violate one or more} Bell/CHSH inequalities cannot be replicated by such instruction sets or by a manipulative Charlie, and so provide evidence that Alice and Bob are jointly observing a single objective system $\mathbb{S}$.  Such correlations can, in particular, identify an entangled state of $\mathbb{S}$ that Alice and Bob share {(for recent experimental demonstrations, see \cite{hensen:15,giustina:15,shalm:15})}.  They cannot, however, by Theorem~1, specify the complete state space of $\mathbb{S}$.  Therefore, they cannot identify the one system that Alice and Bob are guaranteed to share under any circumstances, viz. the system comprising everything in the universe except Alice and Bob.

Two features of the use of Bell/CHSH inequalities as an entanglement witness are of particular relevance to system identification.  First, at least one of the $A_i$ and one of the $B_i$ must measure the state of a pointer observable not manipulable by Charlie.  In the canonical EPR-type experimental setup, these observables correspond to the orientation settings for the polarization/spin measurements, which are assumed to be freely chosen by Alice and Bob, respectively, at each $t$.  This free-choice assumption rules out super-determinism \cite{hofer-szabo:17} by preventing Charlie from specifying correlations involving these settings.  As seen in Section~\ref{leggett} above, the existence of at least one pointer observable controlled by the observer enables objective system identification over time.  Hence, the free-choice assumption can also be viewed as the assumption that Alice and Bob can each, independently, identify their respective apparatus as objective.  Second, Alice and Bob must, after their observations have been completed, exchange a classical message encoding their observational outcomes to compute the correlations observed.  This separate, classical communication step  (i.e., use of a LOCC (Local Operations, Classical Communication) 
  protocol) is required for shared entanglement to serve as a communication resource \cite{bartlett:07}.  It introduces a second system, the classical message, that Alice and Bob must share, but without the restriction of spacelike separation.  The~separate, local observations employed in a LOCC protocol can be regarded as detecting a Bell/CHSH inequality violation only if the joint identification of this later, classical message---in practice, 
Alice and Bob agreeing that they have securely shared reports of their outcomes---is regarded as unproblematic.

Theorem~3 shows that joint system identification by spacelike-separated observers is demonstrable empirically only within quantum theory; in classical theory it can at best be \textit{assumed}.  By showing that joint system identification requires use of a LOCC protocol, it suggests that all systems are equivalent to communication channels.  This idea is implicit in operational reconstructions of quantum theory~\cite{coecke:10,chiribella:11} and has been made explicit by Grinbaum \cite{grinbaum:17}.


\vspace{6pt} 



\textbf{Acknowledgements:}  This research was funded by the Federico and Elvia Faggin Foundation.  Thanks to Mauro D'Ariano, Don Hoffman, Ken Krechmer, Antonino Marcian\`{o}, Chetan Prakash, Robert Prentner and participants in the Quantum Contextuality in Quantum Mechanics and Beyond 2018 workshop for relevant discussions, and to three anonymous referees and the academic editor for helpful comments.

~\\
The author declares no conflict of interest.  The funding sponsor had no role in the design of the study, the writing of the manuscript, or the decision to publish the results.

The following abbreviations are used in this manuscript:\\

\noindent 
\begin{tabular}{@{}ll}
ADC & Analog-to-Digital Converter \\
CHSH & Clauser-Horne-Shimony-Holt \\
EPR & Einstein-Podolsky-Rosen \\
LOCC & Local Operations, Classical Communication \\
POVM & Positive Operator-Valued Measure
\end{tabular}


\end{document}